\documentclass[12pt,aps]{revtex4}

\usepackage{graphicx}
\usepackage{bbm}
\usepackage{amssymb}
\usepackage{amsmath}

\newcommand{\nc}{\newcommand}
\nc{\be}{\begin{equation}}
\nc{\ee}{\end{equation}}
\nc{\bea}{\begin{eqnarray}}
\nc{\eea}{\end{eqnarray}}
\nc{\nn}{\nonumber}
     
\nc{\rref}{Ref.}
\nc{\rrefs}{Refs.}
\nc{\simlt}{\stackrel{<}{{}_\sim}}
\nc{\simgt}{\stackrel{>}{{}_\sim}}
\nc{\eq}{Eq.}
\nc{\eqs}{Eqs.}

\begin{document}

\begin{flushright}
CERN-PH-TH/2008-196\\
IFT-UAM/CSIC-08-54\\
UAB-FT-655
\end{flushright}
\vspace{1cm}
\title{ Some Cosmological Implications of Hidden Sectors}

\author{{\bf J.~R.~Espinosa $^a$, T. Konstandin $^b$, 
J.~M.~No $^a$ and M.~Quir\'os $^{b,c,d}$}}

\email[]{jose.espinosa@cern.ch}
\email[]{konstand@ifae.es}
\email[]{josemi.no@uam.es}
\email[]{quiros@ifae.es}

\affiliation{$^a\,$ IFT-UAM/CSIC, Fac. de Ciencias UAM, 
E-28049 Madrid, Spain}

\affiliation{$^b\,$ IFAE, Univ. Aut\`onoma de Barcelona, 
E-08193 Bellaterra, Barcelona, Spain}

\affiliation{$^c\,$ Instituci\`o Catalana de Recerca i Estudis 
Avan\c{c}ats (ICREA)}

\affiliation{$^d\,$ Theory Division, Physics Department, CERN, 
CH 1211 Geneva 23}

\date{\today}

\begin{abstract}
We discuss some cosmological implications of extensions of the
Standard Model with hidden sector scalars coupled to the Higgs
boson. We put special emphasis on the conformal case, in which the
electroweak symmetry is broken radiatively with a Higgs mass above the
experimental limit. Our refined analysis of the electroweak phase
transition in this kind of models strengthens the prediction of a
strongly first-order phase transition as required by electroweak
baryogenesis. We further study gravitational wave production and the
possibility of low-scale inflation as well as a viable dark matter
candidate.
\end{abstract}

\maketitle

%
%%%%%%%%%%%%%%%%%%%%%%%%%%%%%%%%%%%%%%%%%%%%%%%%%%%%%%%%%%%%%%%%%%%%%%%%%%%%%%%
%  MAIN TEXT
%%%%%%%%%%%%%%%%%%%%%%%%%%%%%%%%%%%%%%%%%%%%%%%%%%%%%%%%%%%%%%%%%%%%%%%%%%%%%%%
%

\section{Introduction}

The Standard Model of particle physics (SM) is nowadays considered to
be an effective theory valid only up to a certain physical cutoff
scale. Even though there exist a large variety of extensions of the
SM, models with a hidden sector have recently attracted some
attention. We will consider models with additional scalar fields that
might transform non-trivially under a hidden gauge group but which are
singlets under the SM gauge group. The only renormalizable interaction
of such scalars with the SM occurs via the Higgs sector, which in this
case serves as a portal to the hidden sector~\cite{hidden_sec}.

In this paper we are concerned with some of the possible cosmological
implications of hidden sector extensions of the SM. This is a
continuation of the study of the electroweak breaking and phase
transition presented in Ref.~\cite{Espinosa:2007qk} and we will
provide some technical details that were omitted there. In addition we
will present an analysis of other cosmological implications, namely
gravitational wave production and dark matter abundance. We also
comment on the possibility of low-scale inflation and present a
calculation of the bubble wall velocity in case of a first-order
electroweak phase transition. As in Ref.~\cite{Espinosa:2007qk} we pay
special attention to the classically conformal case which, for a strong
coupling between the hidden sector scalars and the Higgs field, can be
consistent with the mass bounds on the Higgs particle.

The paper is organized as follows. In Section~\ref{sec_model}, the
model is presented, both at zero and finite temperature.  In
Section~\ref{sec_pheno} the cosmological implications of the model
mentioned above are discussed and we conclude in
Section~\ref{sec_concl}.

\section{The Model \label{sec_model}}

\subsection{Zero Temperature Potential}

We consider a set of $N_S$ real scalar fields $S_i$ coupled to
the SM Higgs doublet $H$ with the tree level potential
\be
\label{eq_V0}
 V_0 = m^2 H^\dagger H + \lambda ( H^\dagger H )^2
+ \sum_i  \left( \frac12 m^2_{S_i}
+ \zeta^2_i \, H^\dagger H \right)S_i^2\ .
\ee
We assume there are no linear or cubic terms in the hidden-sector
scalar fields [this can be enforced by some global symmetry,
e.g. $O(N)$]. Besides, we assume that the squared masses of the hidden
scalars are semi-positive definite ($m^2_{S_i}$), such that this
global symmetry remains unbroken and no quartic terms are necessary to
stabilize the potential.

In the presence of a background Higgs field, $\left< H^0 \right> =
\phi /\sqrt{2}$, the one-loop effective potential in Landau gauge and
$\overline{MS}$ scheme is then given by
\be
\label{eq_V1_MS}
 V_{\rm 1-loop} = V_0 + \Delta V_{\rm 1-loop} = \frac{m^2}2 \phi^2 +
 \frac\lambda4 \phi^4 + \sum_\alpha \frac{N_\alpha M_\alpha^4(\phi)}{64 \pi^2}
\left[ \ln \frac{M_\alpha^2(\phi)}{Q^2}- C_\alpha \right]\ .
\ee
The subscript $\alpha =\{ Z, W, t ,H ,G, S_i \}$ denotes the gauge
bosons ($Z^0$ and $W^\pm$), top quark, Higgs boson, Goldstone bosons
($G^0$ and $G^\pm$) and hidden-sector scalar fields with $N_\alpha =
\{ 3,6,-12,1,3, N_S \}$ while $C_\alpha = 5/6$ for gauge bosons and
$3/2$ for fermions and scalars. The $\phi$-dependent tree level masses
are
\bea
& M^2_{S_i}(\phi) = m^2_{S_i}+\zeta^2_i \phi^2\ , \quad 
M^2_Z(\phi)=\frac{1}{4}( g^2 + 
g^{\prime2}) \phi^2\ , 
\quad M^2_W(\phi)= \frac{1}{4} g^2 \phi^2\ ,& \\
& M^2_t(\phi)= \frac{1}{2} y_t^2 \phi^2\ , \quad M^2_H(\phi) = 3 \lambda 
\phi^2 + 
m^2\ , 
\quad M^2_G(\phi)= \lambda \phi^2 + m^2\ ,&
\eea
where $g$ and $g^\prime$ denote the SM gauge couplings and $y_t$ the
top quark Yukawa coupling.

As it was mentioned in the introduction the case with classical
conformal invariance (i.e. $m^2=0$ and $m_{S_i}^2=0$) is especially
interesting. In this situation all masses are proportional to the
Higgs vacuum expectation value (VEV) and no dimensionful parameters
enter into the tree level potential. However conformal invariance is
broken by loop corrections as can be seen in \eq~(\ref{eq_V1_MS}) by
the occurrence of the renormalization scale $Q$. In this way a mass
scale is introduced via dimensional
transmutation~\cite{Espinosa:2007qk,Coleman:1973jx}. Notice that an important
difference with respect to the pure $\phi^4$-theory is that, in the
interesting region of the parameter space, the loop contributions are
dominated by the hidden-sector scalar (and top) particles. Hence, it
is not mandatory to improve the one-loop potential by renormalization
group techniques (unlike in the $\phi^4$-theory) and moreover the
Goldstone and Higgs one-loop contributions to the potential can be
safely neglected~\cite{Coleman:1973jx, Jackiw:1974cv}. In the
classically conformal case the correct VEV follows from the
minimization condition
\be
 \lambda = -  \sum_\alpha \frac{N_\alpha M^4_\alpha(v)}{16 \pi^2 v^4}
\left[ \ln \frac{ M^2_\alpha(v) }{Q^2} - C_\alpha + \frac12 \right]\ ,
\ee
where $v\simeq 246$ GeV is the observed Higgs VEV of the SM and $Q$
should be chosen near $v$. The potential then reads as
\be
\label{eq_V1_conf}
 V^{\rm conf}_{\rm 1-loop} =  \frac{m_H^2}{8 v^2}\phi^4   
\left[ \ln \frac{\phi^2}{v^2} - \frac12 \right] \ ,
\ee
where $m_H$ is the one-loop Higgs mass given by
\be
\label{eq_mHiggs_conformal}
m_H^2 = \left. \frac{\partial^2}{\partial \phi^2} V \right|_{\phi=v}
= \sum_\alpha \frac{N_\alpha M_\alpha^4(v)}{8 \pi^2 v^2}\ .
\ee
One can see that the occurrence of a sizable number of hidden-sector 
scalars, rather strongly coupled to the Higgs field, can lead to 
a Higgs mass above the LEP bound, even if the theory is classically
conformal invariant~\footnote{In this case it is obvious that the
one-loop Higgs mass differs significantly from its tree level value
$m_H^2=3\lambda v^2$. Similarly the Goldstone boson is massless at
one-loop unlike the tree level result $m^2_G= \lambda v^2$.  This
discrepancy between the tree-level and one-loop masses does also
appear in the non-conformal case. It usually does not constitute a
significant complication, since the loop contributions of the Higgs
and Goldstone bosons are small.}. 

Given the fact that the dramatic impact on electroweak symmetry
breaking we find is due to a sizable number of scalars somewhat
strongly coupled to the Higgs, one might worry about the stability of
the results when higher-order corrections to the potential are
included. It is straightforward to obtain the dominant two-loop
radiative corrections to the Higgs potential (those that depend on the
top Yukawa coupling $y_t$ and $\zeta$) by using standard techniques, as
e.g.~those used in Ref.~\cite{EZ}.  We have found that these two-loop
effects never modify the structure of the potential in a qualitative
way.

Finally, we would like to comment on the influence of the hidden-sector
scalars on the cubic Higgs self-coupling. In \rref~\cite{Noble:2007kk}
the claim was made that a strong phase transition often would lead to
a deviation of the cubic Higgs coupling from its SM value. Taking only
into account the top and hidden-sector scalar contributions, one
obtains
\be
\label{eq_cubic_dev}
\frac{\partial_\phi^3 V}{\partial_\phi^3 V^{\rm SM}} -1
= \frac{\sum_i \zeta_i ^4}{ 12 \pi^2 M_H^2 / v^2 - 3 y_t^4 }\ ,
\ee
which will be correlated with the strength of the phase transition in a
later section.

\subsection{Finite Temperature Potential}

In order to study the electroweak phase transition of the model, we
consider the one-loop potential at finite temperature including the
resummed Daisy-diagrams. The corresponding contributions are given by
\bea
\label{eq_pot_thermal}
\Delta V_T &=& \frac{T^4}{2 \pi^2} \sum_\alpha N_\alpha \int dx \,  x^2 \,
\log[ 1 \pm \exp(- \sqrt{x^2 + M^2_\alpha/T^2})] \nonumber\\ 
&& + \frac{T}{12\pi} \sum_{\alpha \in \textrm{bosons}} N_\alpha 
\left\{ M^3_\alpha  -[M^2_\alpha + \Pi_\alpha(T)]^{3/2} \right\}\ ,
\eea
where the $+ (-)$ holds for fermions (bosons) and $\Pi_\alpha(T)$ are
the thermal masses of the different bosonic species. Neglecting small
$g'$ contributions they read
\bea
\Pi_G&=& \Pi_H = \left( \frac12 g^2 +   \frac14 y_t^2
+\frac12 \lambda + \frac1{12} \sum_i \zeta_i^2 \right) T^2, \\
\Pi_{S_i} &=& \frac13 \zeta_i^2 T^2, \quad 
\Pi_W =  \Pi_Z = \frac{11}6 g^2 T^2\ .
\eea
Besides, in the resummed Daisy diagrams only the longitudinal
polarizations of the gauge bosons contribute.

\section{Cosmology of Hidden-Sector Scalars \label{sec_pheno}}

In this section we discuss cosmological implications of the
hidden-sector scalar extensions of the SM. Namely, the electroweak
phase transition, low-scale inflation, the bubble wall velocity during
a first-order phase transition, gravitational wave production and dark
matter are analyzed.
%We also comment on electroweak baryogenesis.

\subsection{Electroweak Phase Transition}

In order to study the electroweak phase transition, we determine the
so-called bounce solution of the three-dimensional Euclidean action
that quantifies the tunneling probability in case of a first-order
phase transition~\cite{Coleman:1977py, Callan:1977pt, Linde:1980tt}.

At finite temperature the bounce solution is obtained by extremizing
the action
\be
S_3 = 4\pi \int_0^\infty d\rho\ \rho^2 
\left[ \left( \frac{d\phi}{d\rho} \right)^2 + V(\phi) \right]\ ,
\ee
(where $\rho$ is the radial distance from the center of the bubble) with 
solutions obeying the boundary conditions
\be
\partial_\rho \phi (0) = 0, \quad
\lim_{\rho \to \infty} \phi(\rho) = 0\ .
\ee
In addition it is understood that the bounce solution $\phi$ starts
(at $\rho=0$) close to the global minimum of the potential (the broken
phase of the Higgs vacuum).

The tunneling rate per unit volume and time element is approximately
given by \cite{Linde:1980tt}
\be
\label{eq_rate_3d}
\Gamma \simeq \kappa_3\, T^4 \, \exp [- 
S_3(T)/T ]\ ,
\ee
with $\kappa_3=[S_3(T)/(2\pi T)]^{3/2}$, such that the average number of 
bubble nucleations per Hubble volume is given by
\be
\label{eq_P_def}
P(T) =  \int^{T_c}_T  \kappa_3 \frac{d\tilde T}{\tilde T} 
\frac{\tilde T^4}{H^4}   
\exp[- S_3(\tilde T)/ \tilde T ]\ ,
\ee
where the Hubble parameter is given by
\be
\label{eq_def_Hubble}
H^2 \simeq \frac{8 \pi^3 g_* T^4}{90 M^2_{Pl}}\ ,
\ee
$g_*\simeq 106.75 + N_S$ is the effective number of degrees of
freedom and $M_{Pl} = 1.22 \times 10^{19}$~GeV is the Planck mass.

Tunneling becomes in principle possible below the temperature $T_c$ at
which the two minima of the potential are degenerate, but for almost
degenerate vacua, the tunneling rate is still too small to start the
phase transition. We define the temperature $T_n$ at which the phase
transition starts by the average occurrence of one bubble per Hubble
volume
\be
\label{eq_Tn_def}
P(T_n) = 1.
\ee
The first nucleation of bubbles will hence approximately take place
when
\be
\label{eq_140}
\frac{S_3(T_n)}{T_n} \simeq 4 \log \left( \frac{T_n}{H} \right) 
\simeq 142  - 4 \log \left( \frac{T_n}{v} \right) .
\ee

In order to characterize the end of the phase transition the fraction
of space that is covered by bubbles can be used. Neglecting overlapping 
bubbles this is given by
\be
f (T) = \frac{4\pi}{3} \int^{T_c}_T \kappa_3\, \frac{d\tilde T}{\tilde T} 
\frac{\tilde T^4}{H}   
\, R^3(T,\tilde T) \, \exp[- S_3(\tilde T)/ \tilde T]\ , \quad
\ee
where
\be
R(T, \tilde T) = \frac{v_b}{H} \left( 1- \frac{T}{\tilde T} \right)\ .
\ee
Here $v_b\simeq 1$ is the velocity of the bubble wall and we define
the end of the phase transition $T_f$ by
\be
f(T_f) = 1\ .
\ee
\begin{figure}
\begin{center}
\includegraphics*[width=0.7 \textwidth]{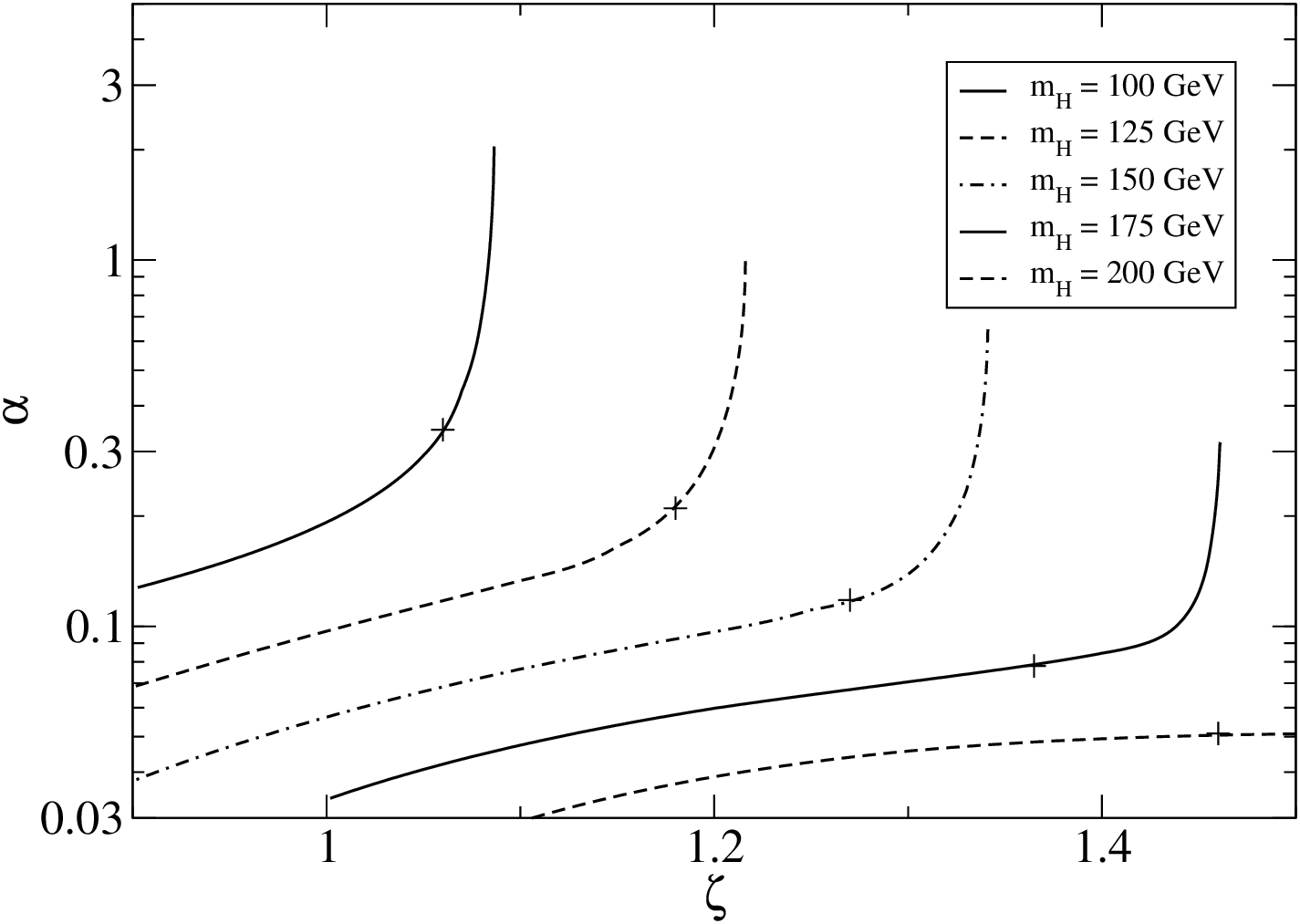}
\\
\vspace*{0.2cm}
\includegraphics*[width=0.73 \textwidth]{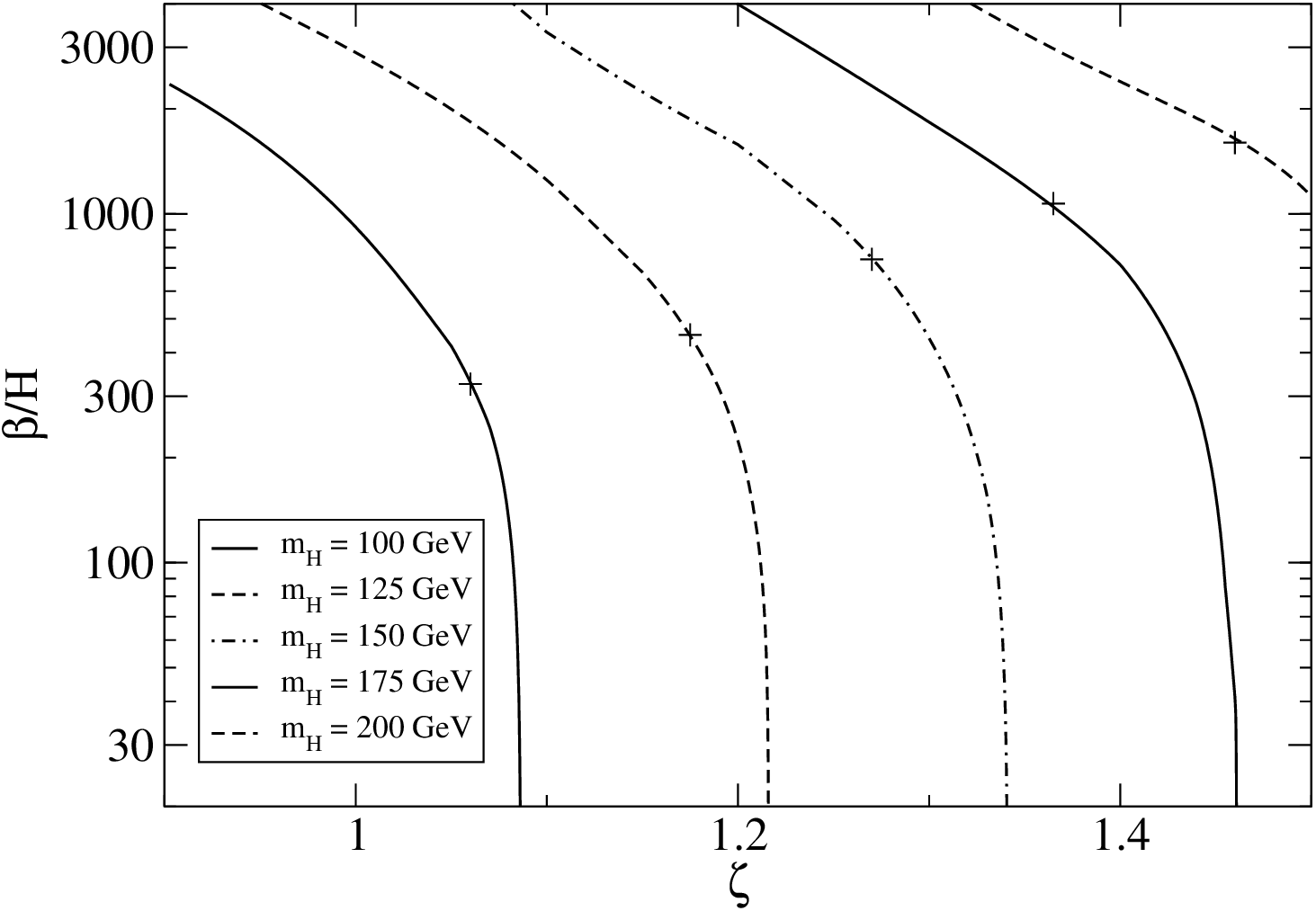}
\end{center}
%\vskip -1cm
\caption{
The parameters $\alpha$ and $\beta$ characterizing the electroweak
phase transition as functions of $\zeta$ for several Higgs masses. A
universal coupling $\zeta$ and $N_S=12$ scalar fields have been
used. The crosses mark the conformal case. }
\label{fig_PTab}
\end{figure}
\begin{figure}
\begin{center}
\includegraphics*[width=0.7 \textwidth]{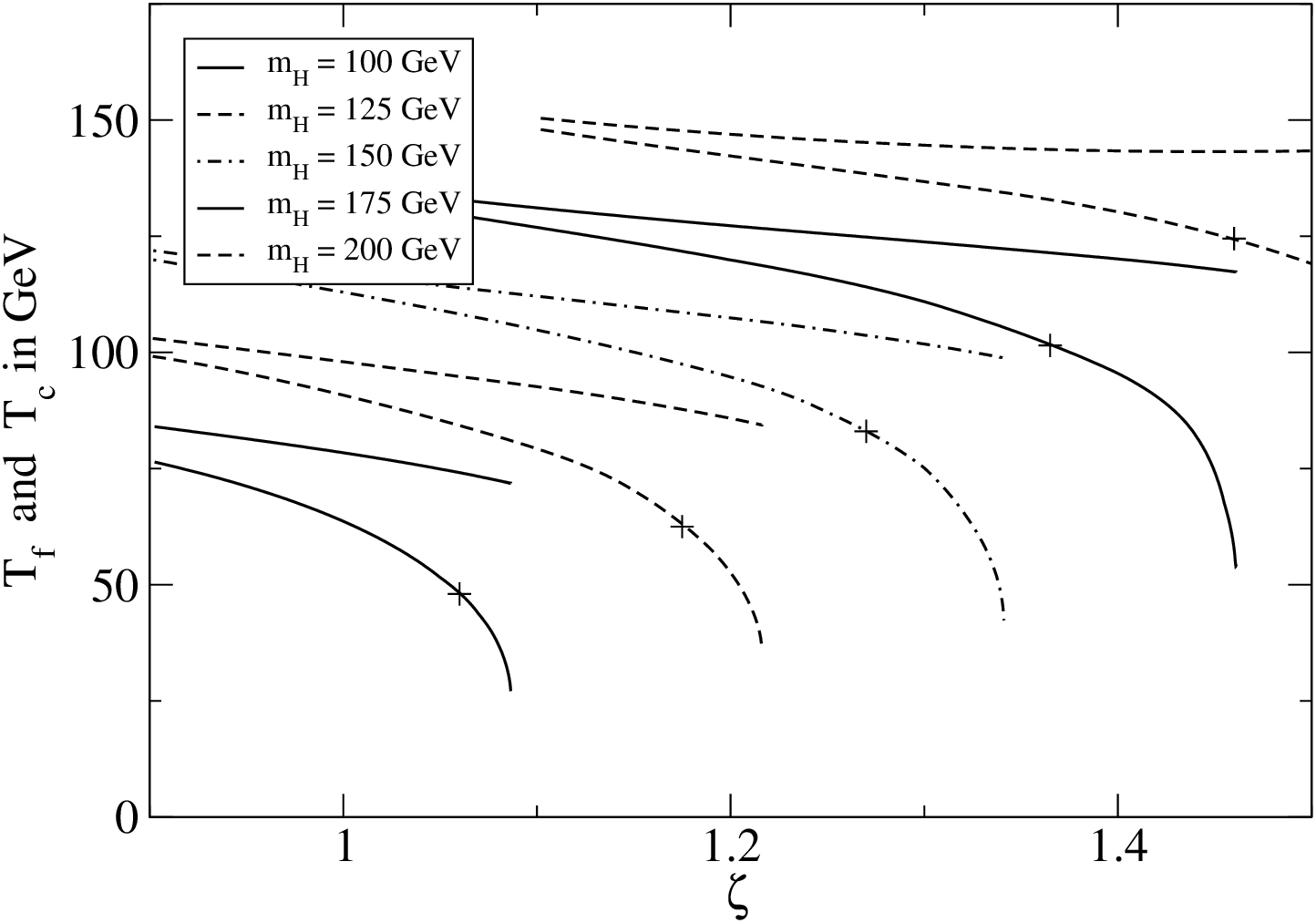}
\end{center}
%\vskip -1cm
\caption{ Same as in Fig.~\ref{fig_PTab} but for the critical
temperature for vacuum degeneracy, $T_c$ (upper curve), and the
temperature at the end of the phase transition, $T_f$ (lower curve). }
\label{fig_PTtemps}
\end{figure}
\begin{figure}
\begin{center}
\includegraphics*[width=0.7 \textwidth]{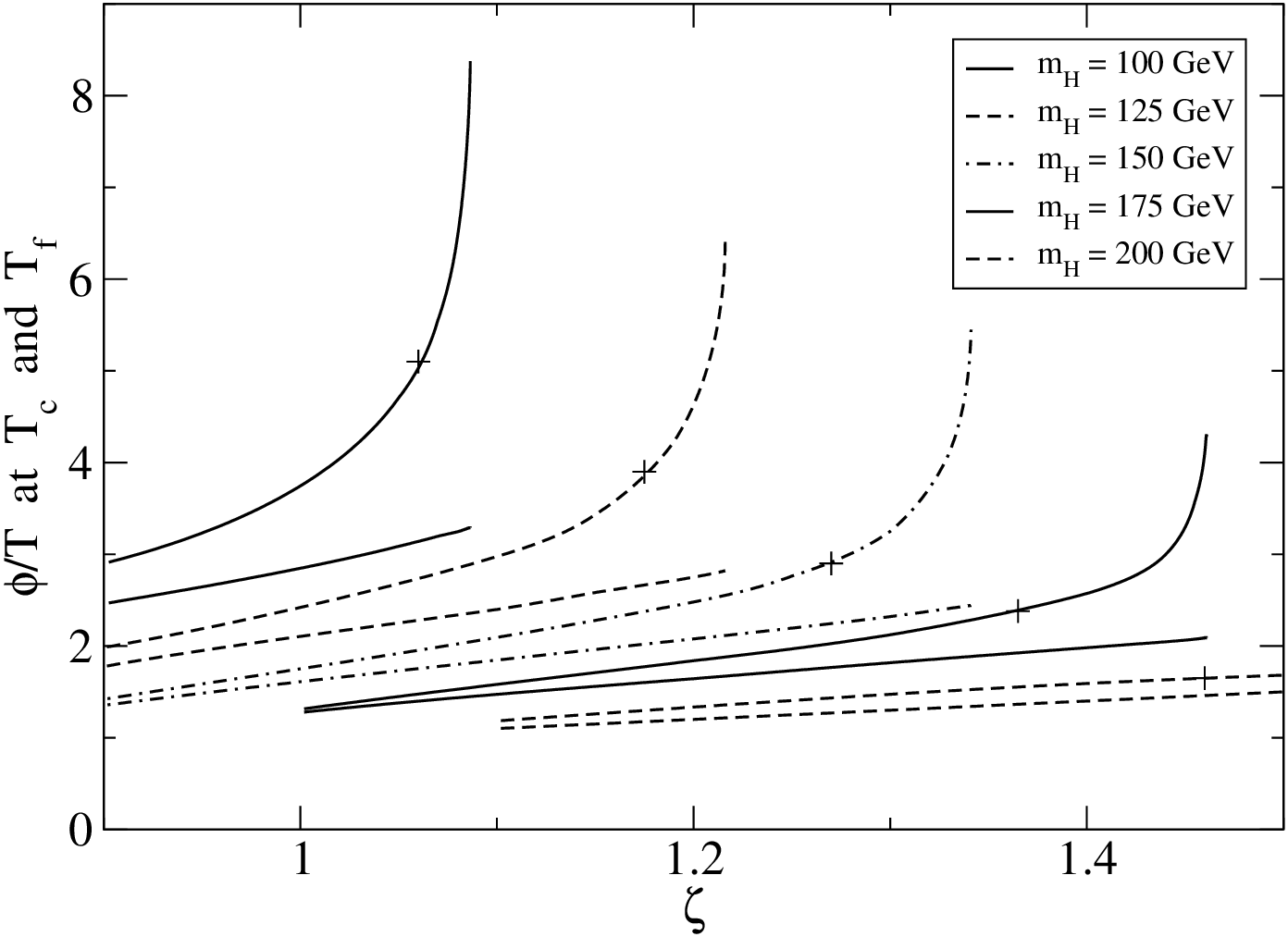}
\end{center}
%\vskip -1cm
\caption{
Same as in Fig.~\ref{fig_PTab} for the ratio $\phi/T$ at the critical
temperature $T_c$ (lower curve) and at the end of the phase
transition, when the temperature is $T_f$ (upper curve). }
\label{fig_PTphioT}
\end{figure}
\begin{figure}
\begin{center}
\includegraphics*[width=0.7 \textwidth]{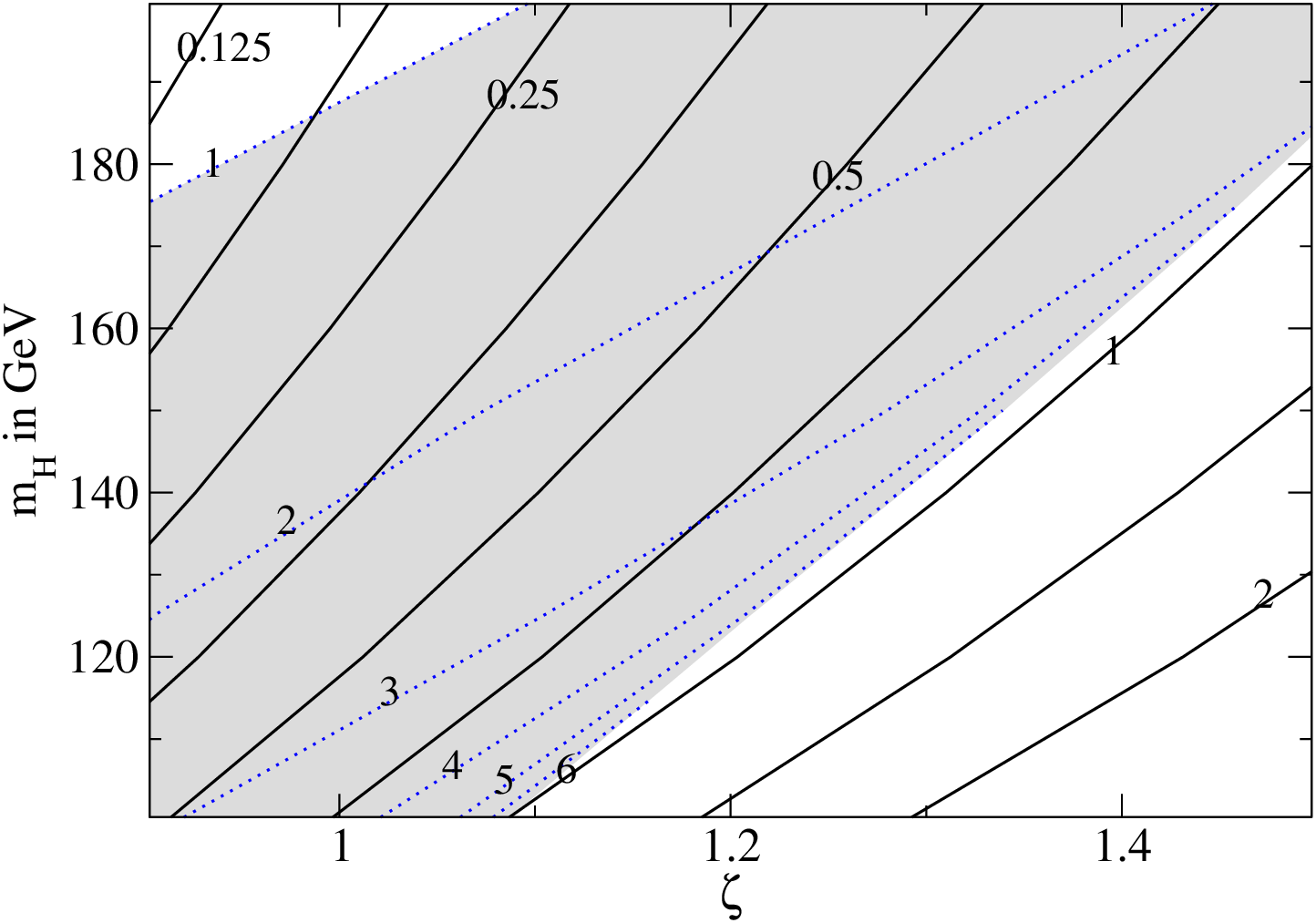}
\end{center}
%\vskip -1cm
\caption{
The deviation of the cubic Higgs coupling from its value in the
Standard Model (solid lines). The shaded region corresponds to a
strong first-order phase transition: the dotted lines are labeled by
the corresponding value of $\phi/T_f$.}
\label{fig_PTVcub}
\end{figure}

In order to quantify the strength of the phase transition we determine
several quantities. These are evaluated at the end of the phase
transition, when most cosmological processes such as baryogenesis and
gravitational wave production take place. The first quantity is the
ratio between the Higgs VEV and the temperature, $\phi(T) / T$. This
ratio is important for baryogenesis, since suppression of washout
effects by sphalerons~\cite{Farrar:1993hn} requires $\phi(T) / T
\gtrsim 1.0$ in the Standard Model.  We do not expect this bound to be
much different in the present model, since the sphaleron energy is
dominated by the contributions from the gauge field configurations
excited in the sphaleron rather than the scalar
ones~\cite{Klinkhamer:1984di}. The second quantity is the duration of
the phase transition $1/\beta$, which is given by
\be
\frac{\beta}{H} = T \frac{d}{dT} 
\left( \frac{S_3}{T}\right)\ .
\ee
The last quantity we are interested in is the latent heat
\be
\epsilon = T \frac{d(V(\phi)-V(0))}{dT} - V(\phi) + V(0)\ .
\ee
The latent heat is usually normalized to the energy density of the
radiation in the plasma, through the dimensionless parameter
$\alpha$
\be
\label{eq_def_alpha}
\alpha = \frac{\epsilon}{ \rho_{\rm rad}} = \frac{30 \epsilon }{\pi^2 g_* 
T^4}\ .
\ee
The quantities $\alpha$ and $\beta$, as well as the bubble velocity
$v_b$, are the key parameters that govern gravitational wave
production (discussed in a later section).

For our numerical examples we take, as in
\rref~\cite{Espinosa:2007qk}, a number of scalars $N_S=12$ with
universal couplings to the Higgs, $\zeta_i = \zeta$, and no explicit
mass terms, $m_{S_i}=0$. The results for the electroweak phase
transition parameters listed above, as functions of $\zeta$ and for
several values of the Higgs mass $m_H$ (consistent with electroweak
breaking conditions), are plotted in Figs.~\ref{fig_PTab} to
\ref{fig_PTphioT}.  For small values of $\zeta$, the phase
transition is SM-like and therefore it is of second-order or 
a cross-over. As expected, the phase transition is in general stronger for
larger values of $\zeta$ and smaller Higgs masses. The latent heat (as
described by $\alpha$) and the strength of the transition [as measured
by $\phi(T)/T$] are both quickly increasing with $\zeta$ and larger
for smaller $m_H$ (see Fig.~\ref{fig_PTab}). In the figures we mark 
the conformal case with a cross and we
see that, even in this case, the model shows a first-order phase
transition strong enough to allow for electroweak baryogenesis 
(see Fig.~\ref{fig_PTphioT}). To the
right of that conformal point the Higgs potential of the model has a
barrier separating the symmetric and broken phases even at $T=0$. For
too large values of $\zeta$ this barrier becomes too high and
tunneling by thermal fluctuations is not efficient to trigger the
electroweak phase transition.  Note how the time of the transition,
$1/\beta$, gets larger and larger with increasing $\zeta$. Eventually
no thermal transition will occur beyond a critical point $\zeta_c$ and
one would get stuck in the symmetric minimum (see below).

Finally we point out that in the present model a strong first-order
phase transition does not necessarily imply a very large deviation of
the cubic Higgs coupling from its SM value. Independently of the value
of the Higgs mass, a phase transition that is strong enough for the
suppression of sphaleron processes, $\phi(T)/T \gtrsim 1.0$, is
possible for deviations of the cubic coupling as small as $15\%$, as
can be seen from Fig.~\ref{fig_PTVcub}.

\subsection{Low-Scale Inflation}

Every time a relatively strong first-order phase transition occurs
during the evolution of the Universe the plasma undergoes a stage of
large over-cooling. This means that the energy finally released as
latent heat is large compared to the thermal energy stored in the
plasma. In this context it is worthwhile to ask whether during the
stage of over-cooling the expansion of the Universe is significantly
accelerated due to the dominance of the vacuum energy, {\it i.e.}
whether inflation occurs. This might be interesting in order to
connect the predictions of inflation to low energy physics, but also
for more exotic scenarios like cold electroweak
baryogenesis~\cite{GarciaBellido:1999sv,Tranberg:2003gi}.

However if inflation takes place at electroweak scales the problem on
whether this scenario allows for a graceful exit arises. It is well
known that a realistic first-order phase transition cannot proceed
arbitrarily slow, since percolation requires the decay rate of the
vacuum to become rather large at a certain
temperature~\cite{Guth:1982pn}. This severely constrains the prospects
of low-scale inflation in such scenarios. We will now analyze the two
possible scenarios.

The first scenario is that inflation ends by thermal
tunneling~\cite{Linde:1980tt}. In this case a substantial amount of
inflation is hardly achieved as can be seen as follows. Suppose that
nucleation takes place after $N_e$ e-folds of inflation. In this case
the nucleation temperature, $T_n$, will be very low compared to the
temperature of degenerate vacua $T_c\simlt v$
\be
N_e = \log{(T_c/T_n)} \lesssim \log{(v /T_n)}\ .
\ee
The Higgs VEV will have to remain stuck in the symmetric phase down to 
very small temperatures and the energy density will be dominated by
the vacuum energy, $V(0)$, rather than by the thermal energy of the 
plasma. Roughly speaking, the vacuum energy is related to the
temperature of degenerate vacua, such that the Hubble parameter is
\be
H^2 \propto \frac{V(0)}{M^2_{Pl}}\propto\frac{v^2T_c^2}{M^2_{Pl}}\ . 
\ee
Imposing that the thermal decay rate at $T_n$ is larger than the 
Hubble rate we get the condition
\be
\frac{S_3(T_n)}{T_n} \lesssim 4 \log \left( \frac{T_n}{H} \right) 
\simeq 142  + 4 \log \left( \frac{ T_n }{T_c} \right) 
=  142 - 4 N_e \lesssim 142 \ . 
\ee
In this regime of very small nucleation temperature the
three-dimensional action (that increases with temperature) has
therefore to be much smaller than the electroweak scale:
\be
S_3(T_n\ll v) \lesssim 142 \, T_n \simeq 142 \, T_c \,\, e^{-N_e}  \ll v\ 
.
\ee
This requires the potential barrier at zero temperature to be very
small and we are thus led to the parameter region near the conformal
case (in the conformal limit the barrier and the three-dimensional
tunnel action vanish). In particular, the parameter $\zeta$ cannot be
much larger than in the conformal case. However we know that near the
conformal case the Higgs VEV does not get stuck at the origin: in fact
the phase transition occurs already at temperatures of electroweak
size, as can be seen in Fig.~\ref{fig_PTtemps}.

The fact that $T_n\simlt T_c$ in the conformal case can be understood as 
follows: The potential difference between the symmetric minimum and the 
broken minimum is in this case given by
\be
\label{eq_Vdiff}
V(v) \simeq \frac{m_H^2 v^2}{16}\ .
\ee
The comparison of Eq.~(\ref{eq_Vdiff}) with the thermal contributions
to the potential in Eq.~(\ref{eq_pot_thermal}) shows that the
temperature $T_c$ where the two minima are degenerate is of order of
the Higgs mass.  At the same time, the potential barrier between the
minima is absent at zero temperature, arising solely by temperature
effects. Therefore one expects that the phase transition takes place
with a nucleation temperature $T_n$ of order $T_c$. In particular, the
temperatures $T_n$ and $T_c$ increase with increasing Higgs mass. All
this agrees with the numerical results as shown in
Fig.~\ref{fig_PTtemps}.

The second scenario is that the minimum at the origin does not decay
by thermal fluctuations but rather through vacuum (quantum)
tunneling~\cite{Coleman:1977py}. Instead of \eq~(\ref{eq_rate_3d}),
the tunneling rate for vacuum decay is given by
\be
\label{eq_rate_4d}
\Gamma \simeq \kappa_4 \, v^4 \, \exp[- S_4(T)]\ ,
\ee
where $S_4$ denotes the four-dimensional tunnel action~\footnote{The
four-dimensional bounce solution can be trusted as long as its radius
is smaller than $1/T$. Above some (small) temperature the
three-dimensional bounce solution should be considered instead.}  and
$\kappa_4 =[S_4 /(2\pi)]^2$. Also in this case the first-order phase
transition should not proceed arbitrarily slow and percolation requires,
similarly to Eq.~(\ref{eq_140}),
\be
S_4(T_n) \simeq 4 \log \left( \frac{v}{H}\right)
\simeq 142 - 8 \log \left( \frac{v}{T_n}\right)\ . 
\ee
In order to assure that the minimum at the origin does not decay by
thermal fluctuations, we should be in a parameter region in which
there is a sufficiently large barrier at zero $T$. This occurs when
$\zeta$ is larger than its conformal value that can be read off (for a
fixed Higgs mass) from Fig.~\ref{fig_PTab}. Numerically, one finds
that in this large-barrier regime the quantum tunneling probability is
suppressed ($S_4 \gtrsim 200$), such that quantum tunneling cannot
provide a graceful exit.

\begin{figure}
\begin{center}
\includegraphics*[width=0.7 \textwidth]{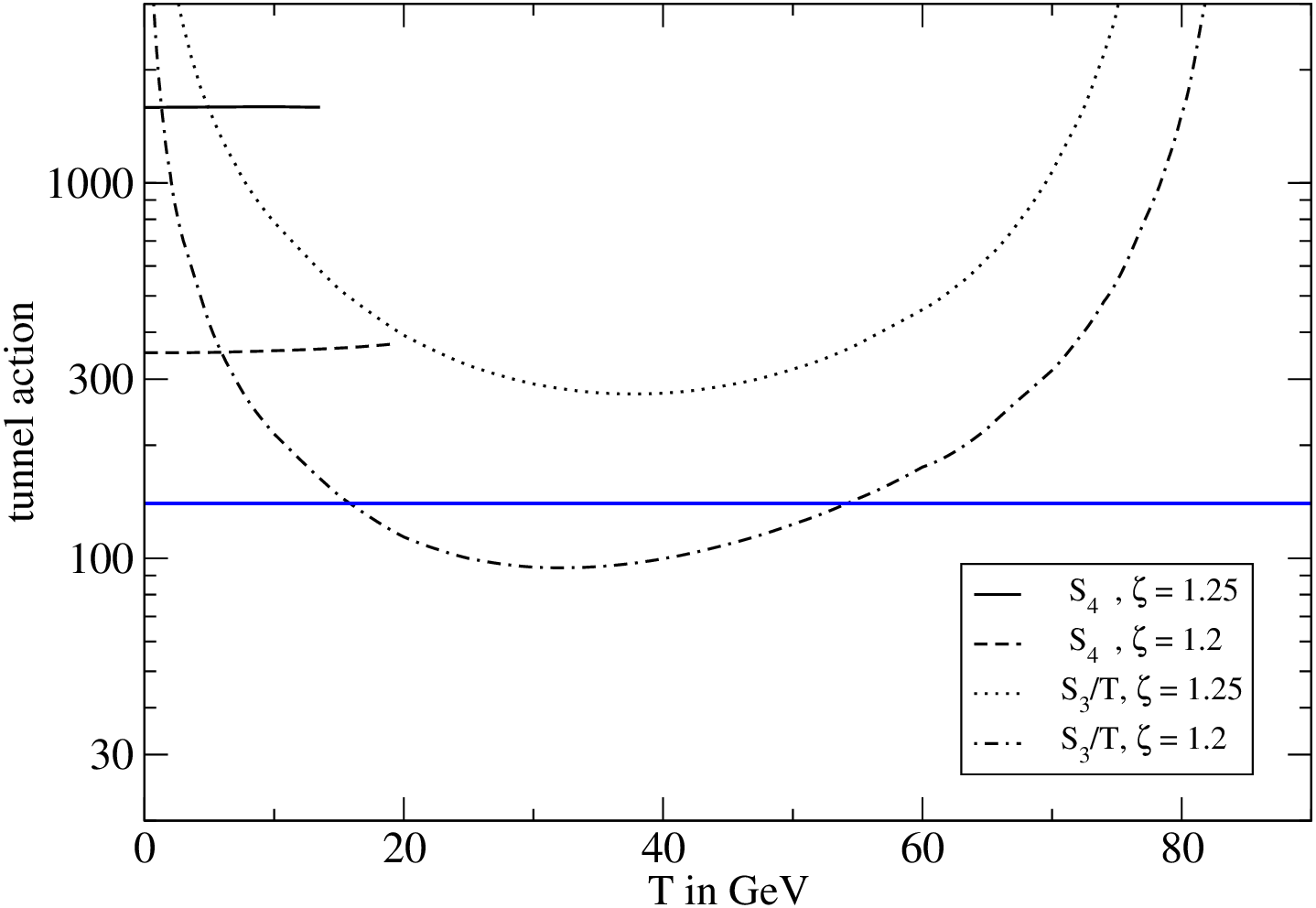}
\end{center}
%\vskip -1cm
\caption{Tunneling actions, $S_3/T$ and $S_4$, as a function of 
temperature for $M_H=125$ GeV and two different values of the coupling
$\zeta$ as indicated. The curves for $S_4$ are stopped when the
four-dimensional bounce ceases to be reliable.}
\label{fig:actions}
\end{figure}

This result was to be expected, since to obtain stability under
thermal fluctuations, a barrier comparable to the difference in vacuum
energy is mandatory. In such regime the tunneling actions can be
calculated using the thin-wall approximation and one finds the scaling
behavior
\be
S_3(T) \propto v\left[\frac{v^4}{V(T,v)}\right]^2, 
\quad S_4(T) \propto \left[\frac{v^4}{V(T,v)}\right]^3\ .
\ee
This shows that, for $V(v) \simeq v^4$, quantum tunneling is typically 
as unlikely as tunneling by thermal fluctuations.

As a conclusion we see that, in general, low-scale inflation with only 
one field seems to require two amply separate scales as for example in 
the case discussed in~\rref~\cite{Nardini:2007me}.

To end this section we show in Fig.~\ref{fig:actions} the typical
behavior of the tunneling actions, $S_3/T$ and $S_4$, as functions of
the temperature for $M_H=125$ GeV and two choices of $\zeta$. For
$\zeta=1.2$, $S_3/T$ gets eventually below the critical nucleation
value $\sim 142$ (horizontal line), and the electroweak phase
transition takes place. For $\zeta=1.25$ no satisfactory transition
would occur as the strong suppression will hinder percolation.

\subsection{Bubble Wall Velocities}

During a strongly first-order phase transition the wall velocity of
the expanding nucleated bubbles is an important parameter. For
example, the standard picture of electroweak baryogenesis is based on
the diffusion of charge-asymmetries into the unbroken phase in front
of the wall. This effect is strongly suppressed if the wall expansion
is supersonic, making electroweak baryogenesis implausible. On the
other hand, gravitational wave production requires large wall
velocities. Hence, baryogenesis and gravitational wave production at
electroweak scales seem to be mutually exclusive~\cite{Huber:2007vva}.

The wall velocity does not only depend on the thermo-dynamical
characteristics of the phase transition, but also on the particle
content of the plasma. In particular, bosonic degrees of freedom that
are massless before the phase transition but become heavy due to a
strong coupling to the Higgs VEV exert a strong friction force on the
wall~\cite{Moore:1995ua, Moore:1995si, Moore:2000wx}. In this way the
presence of many hidden-sector scalars leads to subsonic wall
velocities and the phase transition proceeds by {\it deflagration}.
Following the arguments of \rref~\cite{Moore:2000wx}, in the present
case the friction is dominated by the scalars and is given by
\be
\label{eq_def_fric}
\eta \approx \frac{N_S m_D^2 T}{16 \pi L} \log(M_S L)\ ,
\ee
where $ L\approx 1/T$ denotes the thickness of the bubble wall
during the phase transition and $m^2_D = \frac{1}{3} \zeta^2 T^2$ is
the squared Debye mass of the hidden-sector scalars.

The expansion of the bubbles is driven by the pressure produced by the
latent heat, $p=\epsilon/3$. If the friction forces are large, $\eta
\gg p$, the wall velocity can be estimated to be
$v_b=p/\eta$. Comparison with \eq~(\ref{eq_def_fric}) shows that this
is the case if the phase transition is weak, in the sense that $\alpha
\ll (6 \times 10^{-3}) \, \sum_i \zeta_i^2 $.

In the opposite regime, in which friction effects from the plasma on the
bubble wall are negligible, one expects the phase transition to
proceed by {\it detonation}. In this case the bubble wall velocity can
be determined by a self-consistent calculation that leads to
supersonic wall velocities~\cite{Steinhardt:1981ct}. The wall velocity
is then approximately given by
\be
\label{vb}
v_b = \frac{\sqrt{1/3} + \sqrt{\alpha^2 + 2\alpha/3}}{1 + \alpha}\ ,
\ee
where $\alpha$ is the latent heat normalized to the energy density of
the plasma as given in \eq~(\ref{eq_def_alpha}). In
fact, this value for the wall velocity is only an upper bound, since a
larger class of detonation solutions is known to
exist~\cite{Laine:1993ey}, but we use nevertheless this formula in the
analysis of gravitational wave production.

The results presented in Fig.~\ref{fig_PTab} show that in principle
both possibilities can occur in hidden-sector scalar extensions of
the SM without significant tuning.

\subsection{Gravitational Waves}

Another smoking-gun signal of a cosmological first-order phase
transition is gravitational wave (GW) radiation. When the Higgs
bubbles nucleate and expand, a portion of the latent heat is
transformed into kinetic energy of the Higgs field and also into bulk
motion of the plasma that follows the passing bubble wall
profile. When the bubbles finally percolate and collide, this energy
is partially released into gravitational
waves~\cite{Kamionkowski:1993fg, Apreda:2001us, Caprini:2007xq,
Huber:2008hg}. Surprisingly, the only parameters that enter into the
analysis of the gravitational wave radiation by collisions are the
latent heat normalized to the radiation energy $\alpha$, the wall
velocity of the bubbles $v_b$ and the duration of the phase transition
$1 / \beta$.

In principle, there might be additional mechanisms of GW production,
as {\it e.g.} turbulence in the plasma~\cite{GWturbulence}
and/or magnetic fields~\cite{GWmag}. However, for very strong phase
transitions, the peak frequency of the GW spectrum is shifted to lower
frequencies and mostly the high frequency part of the GW spectrum lies
in the range of best experimental sensitivity of the planned space-based 
experiments. The contributions from bubble collisions usually
dominate for $f \gg f_{\rm peak}$, such that at the frequency of best
sensitivity of LISA or BBO, it suffices to consider the contributions
from collisions. A more complete discussion of these issues can be
found in
\rrefs~\cite{Nicolis:2003tg, Grojean:2006bp, Huber:2007vva}.

In the following we summarize the formulas for GW production by bubble
collisions as recently presented in \rref~\cite{Huber:2008hg}. The
peak frequency is given by
\be
f_{\rm peak} \simeq 10.2 \times 10^{-3}\textrm{ mHz }
\left(\frac{\beta}{H}\right)
\left(\frac{T_f}{100 \textrm{ GeV}}\right)\, 
\left(\frac{1.0}{1.8 + v_b^2}\right)\ ,
\ee
whereas the energy density in GWs amounts to 
\be
h^2 \Omega_{\rm peak} = 1.84 \times 10^{-6} \kappa^2 
\frac{H^2}{\beta^2}
\left( \frac{\alpha}{1 + \alpha} \right)^2 \frac{v_b^3}{0.42 + v_b^2}\ .  
\ee
The efficiency factor $\kappa$ indicates the fraction of latent heat
that is transformed into bulk motion of the plasma and finally into
gravitational waves. It is given by~\cite{Kamionkowski:1993fg}
\be
\label{kappa}
\kappa = \frac{1}{1 + 0.715 \alpha} 
\left[ 0.715 \alpha + \frac{4}{27} \sqrt{\frac{3\alpha}{2}}\right]\ .
\ee
The best sensitivity of BBO (LISA) is at $f = 100$ mHz ($f = 1$ mHz)
expected to be $h^2 \Omega = 10^{-17}$ ($h^2 \Omega =
10^{-11}$). Considering that the GW spectrum from collisions scales
approximately as $h^2 \Omega \propto 1/f$ for large
frequencies~\cite{Huber:2008hg}, one obtains for BBO a signal to
sensitivity ratio \be 1.87 \times 10^7 \kappa^2 \left(
\frac{\alpha}{\alpha + 1} \right)^2 \, \left(\frac{T_f}{100 \textrm{
GeV}}\right) \, \left(\frac{H}{\beta}\right) \,
\left(\frac{v_b^3}{0.76 + 2.22 v_b^2 + v_b^4}\right) , 
\ee
and for LISA a value that is smaller by four orders of magnitude. 

Comparison with the parameters of the phase transition in
Fig.~\ref{fig_PTab} shows that, in the present model a signal that is
detectable by BBO is feasible, if the parameter $\zeta$ is rather
close to the critical point $\zeta_c$ beyond which no thermal
tunneling occurs, requiring a tuning in $\zeta$ at the percent
level. For example, the parameters
\be
\alpha = 0.2, \quad \beta/H = 200,\quad T_f = 50 \, \textrm{GeV}, 
\ee
lead to [using (\ref{vb}) and 
(\ref{kappa})] 
\be
v_b = 0.83, \quad \kappa = 0.20,
\ee
and to a signal to sensitivity ratio of $O(10)$. On the other hand,
no observable traces from the electroweak phase transition
are expected at LISA in the present model.

\subsection{Dark Matter}

In this section we investigate if the new scalar degrees of freedom
constitute a viable dark matter candidate. For simplicity we consider
only one hidden-sector scalar as the generalization to several scalars
is straightforward. Singlet dark matter has already been extensively
discussed in the literature (see
e.g.~\rref~\cite{Silveira:1985rk,McDonald:1993ex,
Burgess:2000yq}). Here we focus on two aspects. First, we discuss if
the same scalar species might be responsible both for a strong phase
transition and for dark matter (Ref.~\cite{Hambye:2007vf} addresses
the same question in a different extension of the Higgs sector).
Secondly, we focus on the classically conformal case, in which the
scalar has no explicit mass term.

The scalar has to be stable to constitute a viable dark matter
candidate. This is achieved by the choice of the potential in
\eq~(\ref{eq_V0}), since the scalars are protected from decay by a
$Z_2$ symmetry. In particular we assume that this symmetry is not
spontaneously broken. Nevertheless the scalars annihilate and the
particle density of the scalar obeys the Boltzmann
equation~\cite{McDonald:1993ex}
\be
\frac{dn_S}{dt} = - 3Hn_S - \left< \sigma_{\rm ann} v \right>
(n_S^2 - n^2_{S,eq})\ ,
\ee
where the equilibrium distribution is given by 
\be
n_{S,eq} = T^3 \left(\frac{M_S}{2\pi T}\right)^{3/2} e^{-M_S/T}\ ,
\ee 
and $H$ denotes the Hubble
parameter as given in \eq~(\ref{eq_def_Hubble}). Rescaling the
distribution functions, $f=n/T^3$, one obtains the equation
\be
\frac{df_S}{dT} = \frac{ \left< \sigma_{\rm ann} v \right>}
{ H / T^2} (f_S^2 - f^2_{S,eq})\ .
\ee

The contributions to $\left< \sigma_{\rm ann} v \right>$ from
annihilation to pairs of Higgs-, W-, Z-bosons and SM-fermions are
respectively given by
\bea
\left< \sigma_{\rm ann} v \right> &=& 
\frac{\zeta^4}{16\pi M^2_S} \left( 1- \frac{M_H^2}{M_S^2}\right)^{1/2} 
\left\{1+\frac{3M_H^2}{D_h}(8M_S^2+M_H^2)+\frac{8\zeta^2 
v^2}{D_S}(M_H^2+2\zeta^2 v^2 -2 M_S^2)\right.
\nn \\
&& -\left.\left(\frac{3M_H^2}{D_h}\right)\left(\frac{8\zeta^2
v^2}{D_S}\right)\left[(4M_S^2-M_H^2)(2M_S^2-M_H^2)-M_H M_S 
\Gamma_H\Gamma_S\right]\right\}\nn\\
&& + 
\frac{\zeta^4 M_W^4}{2\pi M_S^2 D_h} 
\left( 1- \frac{M_W^2}{M_S^2}\right)^{1/2} \left[ 2 +  \left( 1 - 
2\frac{M_S^2}{M_W^2}\right)^2\right]
+ \frac{1}{2} (M_W^2\rightarrow M_Z^2)\nn\\
&& + \sum_{\rm fermions} 
\frac{N_f\zeta^4 M_f^2}{\pi D_h} 
\left( 1- \frac{M_f^2}{M_S^2}\right)^{3/2}\ ,
\eea
where 
\bea
D_h&\equiv &  (4 M^2_S - M_H^2)^2 + M_H^2 \Gamma_H^2\ ,\nonumber\\
D_S&\equiv &  (2 M^2_S - M_H^2)^2 + M_S^2 \Gamma_S^2\ ,
\eea
$\Gamma_H \approx 8 \times 10^{-5} M_H$ is the decay width of
the Higgs particle in the SM and $\Gamma_S$ that of the hidden scalars. 
Finally, $N_f=1(3)$ for leptons (quarks).
\begin{figure}
\begin{center}
\includegraphics*[width=\textwidth]{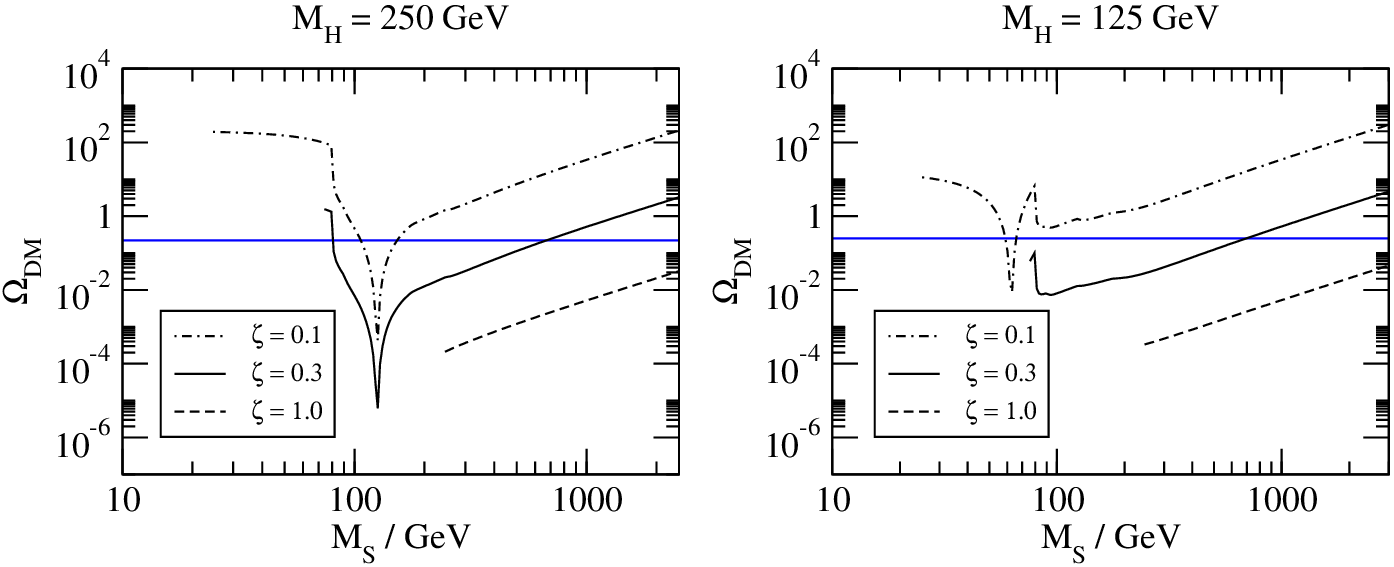}
\end{center}
\vskip -1cm
\caption{
Dark matter density of a single hidden scalar (for two 
different Higgs masses as indicated) as a
function of the scalar mass $M_S$ and different values for its
coupling to the Higgs, $\zeta$.}
\label{fig_DM}
\end{figure}

An approximate solution to this equation was given in
\rref~\cite{Lee:1977ua}. The scalar freeze out temperature $\hat T$
is given by
\be
\frac{M_S}{\hat T} = 
\log\left( \frac{ M_S \left< \sigma_{\rm ann} v \right>} { H / \hat T^2} 
\right)
+\frac12 \log \left( \frac{ 8\pi^3 \hat T } { M_S } \right)\ , 
\ee
and typically one finds $M_S \approx (15 - 25) \, \hat T$. The final
particle density is
\be
f(T \ll M_S) \approx \frac { H / \hat T^2} { \hat T \left< \sigma_{\rm 
ann} v \right>}\ .
\ee
At present, $T=T_\gamma$, the total energy density in scalars is 
\be
\Omega_{DM} = \frac{2}{g_*} \frac{M_S n_S(T_\gamma)}{ \rho_{\rm crit}}
= \frac{2}{g_*} \frac { \, H / T^2} { \hat T \left< \sigma_{\rm ann} v \right>}
\frac{ M_S T_\gamma^3}{ \rho_{crit}}\ ,
\ee
where $\rho_{\rm crit}$ denotes the critical energy density of the
Universe at present.

The dependence of $\Omega_{DM}$ on the scalar mass for fixed coupling
$\zeta$ is plotted in Fig.~\ref{fig_DM}.  Notice that we only plotted
the dark matter density for scalar masses that are larger than $\zeta
v$ and hence correspond to a positive mass term in the Lagrangian
(according to $M^2_S = \zeta^2 v^2 + m^2_S$).

Besides a logarithmic dependence on the freeze-out temperature, the
dark matter density scales for large masses as $ \Omega_{DM} \propto
M_S^2 /\zeta^4$. Notice that for $2M_S \approx M_H$ most annihilation
channels are enhanced and the scalar contribution to dark
matter is suppressed. Finally, the annihilation cross-section drops
considerably below the W-boson threshold, $M_S < M_W$, since if the
scalar is light it mostly annihilates into bottom/anti-bottom pairs,
which is suppressed by the bottom-quark Yukawa coupling. This leads to
an increase of the dark matter density below the W-boson
threshold. Notice that taking temperature effects into account, one
expects that the annihilation cross-section changes less drastically
when the scalar mass is varied. In particular the enhancement close
to the Higgs mass is expected to be less prominent. Likewise, the drop
below the W-boson threshold proceeds in an interval of width $\Delta
M_S \approx T$.

\begin{figure}
\begin{center}
\includegraphics*[width=0.7 \textwidth]{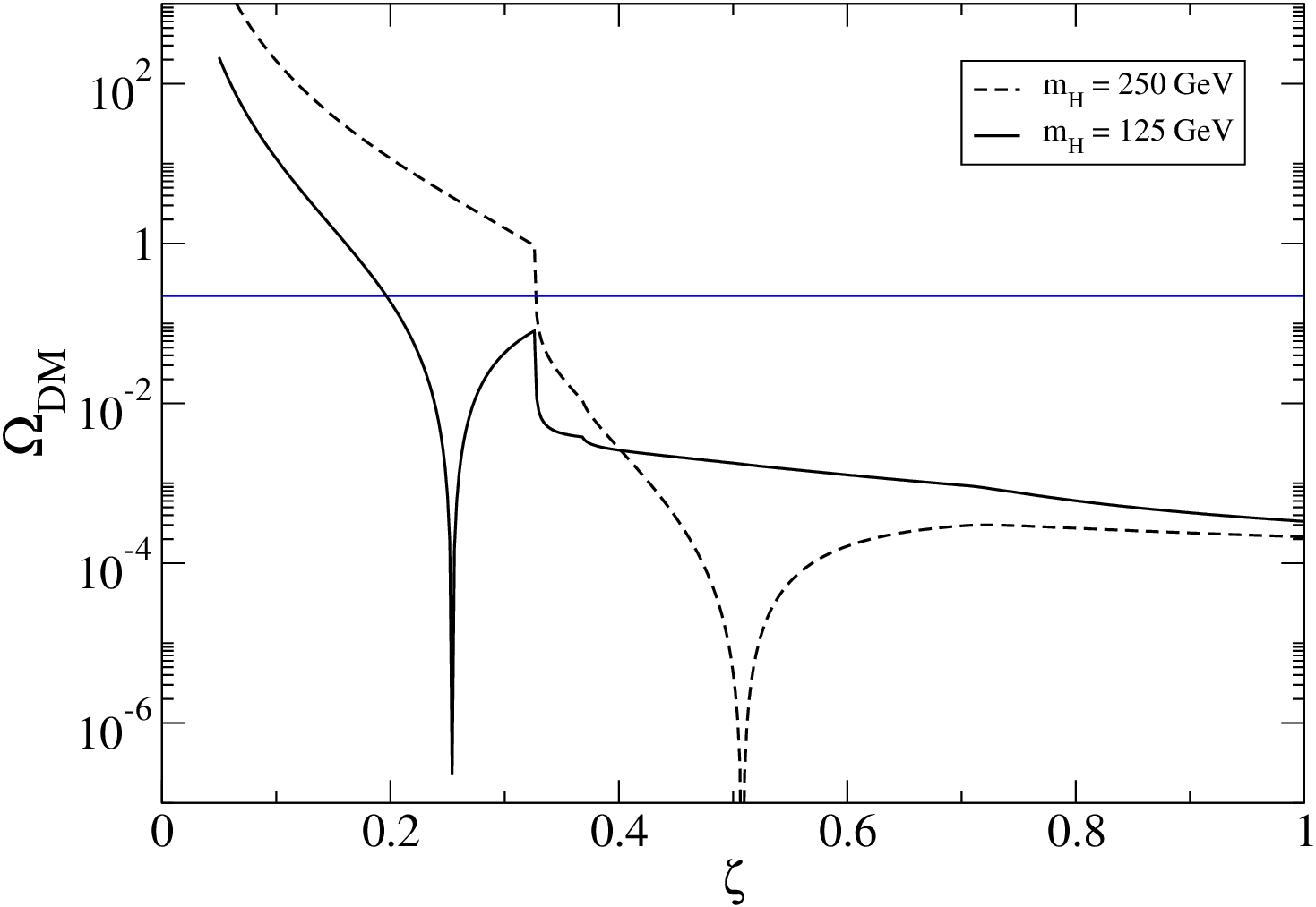}
\end{center}
\vskip -1cm
\caption{ Dark matter density of a single hidden scalar as a function
of the coupling $\zeta$ in the case $M_S=\zeta v$ and for two
different values of the Higgs mass.  }
\label{fig_DMconf}
\end{figure}

Therefore, we see that there are two valid regimes of scalar dark
matter. The first option is to increase the scalar mass term $m_S$,
while keeping the coupling $\zeta$ fixed. However, even in the case of
a rather large number of scalars $N_S=12$, this requires scalar masses
of order TeV and such scalars cannot be responsible for a strong phase
transition. Alternatively, the scalar could be rather light, with $M_S
\lesssim M_W$, and weakly coupled, such that its annihilation is
suppressed. Also in this case, the impact of the scalars on the phase
transition is small.

Finally, consider a model without an explicit singlet mass term 
in the Lagrangian. In Fig.~\ref{fig_DMconf} the dark matter density is
plotted as a function of $\zeta$ for $M_S=\zeta v$ and for two
different values of the Higgs mass. The predicted dark matter density
typically surpasses the observed one below the W-boson
threshold. Besides, in the case that the Higgs boson is lighter than
two W-bosons, the resonant enhancement in the decay channel can lead
to two additional viable values for the parameter $\zeta$ that
reproduce the observed dark matter density. Again, such weakly
coupled scalars cannot increase the strength of the phase transition
considerably. In particular, a classically conformal model requires
several additional, strongly coupled scalars to surpass the current
bounds on the Higgs mass, see Fig. \ref{fig_PTab}.

In conclusion, if extra scalar degrees of freedom are responsible for a
strong electroweak first-order phase transition, as well as for dark matter,
it seems that either the coupling constants $\zeta_i$ or the mass
terms $m_{S,i}$ are non-universal. Scalar dark matter requires either
a scalar with a rather large mass $M_S \approx $ TeV, or a rather
weak coupling $\zeta \approx M_W / v$. However both types of scalars
cannot contribute significantly to the strength of the phase
transition. Hence, the existence of both features in a universal
scalar framework would require a very large number of scalars, which we
estimate to be $N_S \gtrsim 50$.

\section{Conclusions \label{sec_concl}}
 
We have studied several cosmological implications of Standard Model
extensions with hidden-sector scalars. In particular, we strengthen
the results of \cite{Espinosa:2007qk} finding that in
models with a moderate number of hidden-sector scalars, $N_S \approx
12$, the electroweak phase transition is generically of first-order as
long as the Higgs mass is not much larger than the electroweak scale
and the coupling to the hidden sector is substantial, $\zeta \gtrsim
0.9$. An interesting feature of the model is that this property
persists even if the theory is classically conformal invariant and the
electroweak scale is induced by dimensional transmutation. This was
already emphasized in Ref.~\cite{Espinosa:2007qk}. We find
that the phase transition is in a large portion of the parameter space
strong enough to suppress the sphaleron process after the phase
transition, $\phi/T \gtrsim 1.0$ as required by electroweak
baryogenesis. Besides, we find that sizable production of gravitational
radiation requires a tuning of the parameters at the percent level.

Besides a strong first-order phase transition, viable electroweak
baryogenesis requires sizable CP violation. Electroweak baryogenesis
in non-SUSY models typically utilizes a Higgs VEV that has a changing
complex phase during the phase transition. One useful ingredient hence
seems to be to complexify the present scalars and to allow for scalar
VEVs, but still this would not induce a change in the complex phase of
the Higgs VEV such that the introduction of a second Higgs doublet
seems unavoidable. Alternatively, one can introduce an additional
source of CP violation in the quark sector (see
e.g. Ref.~\cite{Bodeker:2004ws}) but undoubtedly CP violation arising
from the hidden sector would be much more appealing in our model.

Concerning dark matter, we find that the abundance required by the concordance
model can be provided by hidden-sector scalars in two different
regimes. In the first, the hidden-sector scalars have moderate
couplings but large masses $M_S \gtrsim 1$ TeV. In the second, the
hidden-sector scalars are rather light, $M_S \lesssim M_W$. In this
case, the scalars cannot annihilate into W-bosons, which greatly
enhances the dark matter abundance. Notice that this scenario is
compatible with scalars that obtain their mass solely by electroweak
symmetry breaking. Nevertheless, neither type of scalar can
contribute significantly to the strength of the phase transition, such
that a viable dark matter candidate cannot substantially improve the
prospects of electroweak baryogenesis compared to the SM. Hence, a
simultaneous solution of the dark matter and baryogenesis problems of
the Standard Model close to electroweak scales either requires a large
number of scalars (in which case we found $N_S \gtrsim 50$), or
several types of scalars in the hidden sector with non-uniform masses
and/or couplings to the Higgs sector.

\section*{Acknowledgments}

J.R.E. thanks A.~Casas, A.~Riotto and G.~Servant for discussions.
Work supported in part by the European Commission under the European
Union through the Marie Curie Research and Training Networks ``Quest
for Unification" (MRTN-CT-2004-503369) and ``UniverseNet"
(MRTN-CT-2006-035863); by a Comunidad de Madrid project (P-ESP-00346);
and by CICYT, Spain, under contracts FPA 2007-60252 and FPA
2005-02211.  T.K. is supported by the EU FP6 Marie Curie Research \&
Training Network 'UniverseNet' (MRTN-CT-2006-035863).


\begin{thebibliography}{99}
\bibliographystyle{unsrt}

%\cite{Schabinger:2005ei}
%\bibitem{Schabinger:2005ei}

\bibitem{hidden_sec}
  R.~Schabinger and J.~D.~Wells,
  ``A minimal spontaneously broken hidden sector and its impact on Higgs  boson
  physics at the Large Hadron Collider,''
  Phys.\ Rev.\  D {\bf 72} (2005) 093007
  [hep-ph/0509209].
  %%CITATION = PHRVA,D72,093007;%%

%\cite{Patt:2006fw}
%\bibitem{Patt:2006fw}
 B.~Patt and F.~Wilczek, 
  ``Higgs-field portal into hidden sectors,''
  [hep-ph/0605188].  %%CITATION = HEP-PH/0605188;%%

%\cite{Bowen:2007ia}
%\bibitem{Bowen:2007ia}
  M.~Bowen, Y.~Cui and J.~D.~Wells,
  ``Narrow trans-TeV Higgs bosons and H $\rightarrow$ h h decays: Two LHC 
search paths
  for a hidden sector Higgs boson,''
  JHEP {\bf 0703} (2007) 036
  [hep-ph/0701035].
  %%CITATION = JHEPA,0703,036;%%

%\cite{Espinosa:2007qk}
\bibitem{Espinosa:2007qk}
  J.~R.~Espinosa and M.~Quir\'os,
  ``Novel effects in electroweak breaking from a hidden sector,''
  Phys.\ Rev.\  D {\bf 76} (2007) 076004
  [hep-ph/0701145].
  %%CITATION = PHRVA,D76,076004;%%


%\cite{Coleman:1973jx}
\bibitem{Coleman:1973jx}
  S.~R.~Coleman and E.~Weinberg,
  ``Radiative Corrections As The Origin Of Spontaneous Symmetry Breaking,''
  Phys.\ Rev.\  D {\bf 7} (1973) 1888.
  %%CITATION = PHRVA,D7,1888;%%

%\cite{Jackiw:1974cv}
\bibitem{Jackiw:1974cv}
  R.~Jackiw,
  ``Functional evaluation of the effective potential,''
  Phys.\ Rev.\  D {\bf 9}, 1686 (1974).
  %%CITATION = PHRVA,D9,1686;%%

%\cite{Espinosa:2000df}
\bibitem{EZ} See e.g.
  J.~R.~Espinosa and R.~J.~Zhang,
``Complete two-loop dominant corrections to the mass of the lightest  
CP-even Higgs boson in the minimal supersymmetric standard model,''
  Nucl.\ Phys.\  B {\bf 586} (2000) 3
  [hep-ph/0003246].
  %%CITATION = NUPHA,B586,3;%%

%\cite{Noble:2007kk}
\bibitem{Noble:2007kk}
  A.~Noble and M.~Perelstein,
  ``Higgs Self-Coupling as a Probe of Electroweak Phase Transition,''
  [hep-ph/0711.3018].
  %%CITATION = ARXIV:0711.3018;%%

%\cite{Coleman:1977py}
\bibitem{Coleman:1977py}
  S.~R.~Coleman,
  ``The Fate Of The False Vacuum. 1. Semiclassical Theory,''
  Phys.\ Rev.\  D {\bf 15}, 2929 (1977)
  [Erratum-ibid.\  D {\bf 16}, 1248 (1977)].
  %%CITATION = PHRVA,D15,2929;%%

%\cite{Callan:1977pt}
\bibitem{Callan:1977pt}
  C.~G.~Callan and S.~R.~Coleman,
  ``The Fate Of The False Vacuum. 2. First Quantum Corrections,''
  Phys.\ Rev.\  D {\bf 16}, 1762 (1977).
  %%CITATION = PHRVA,D16,1762;%%

%\cite{Linde:1980tt}
\bibitem{Linde:1980tt}
  A.~D.~Linde,
  ``Fate Of The False Vacuum At Finite Temperature: Theory And Applications,''
  Phys.\ Lett.\  B {\bf 100}, 37 (1981);
  %%CITATION = PHLTA,B100,37;%%
%\cite{Linde:1981zj}
%\bibitem{Linde:1981zj}
%  A.~D.~Linde,
  ``Decay Of The False Vacuum At Finite Temperature,''
  Nucl.\ Phys.\  B {\bf 216} (1983) 421
  [Erratum-ibid.\  B {\bf 223} (1983) 544].
  %%CITATION = NUPHA,B216,421;%%

%\cite{Farrar:1993hn}
\bibitem{Farrar:1993hn}
  G.~R.~Farrar and M.~E.~Shaposhnikov,
  ``Baryon Asymmetry Of The Universe In The Standard Electroweak Theory,''
  Phys.\ Rev.\  D {\bf 50} (1994) 774
  [hep-ph/9305275].
  %%CITATION = PHRVA,D50,774;%%

%\cite{Klinkhamer:1984di}
\bibitem{Klinkhamer:1984di}
  F.~R.~Klinkhamer and N.~S.~Manton,
  ``A Saddle Point Solution In The Weinberg-Salam Theory,''
  Phys.\ Rev.\  D {\bf 30} (1984) 2212.
  %%CITATION = PHRVA,D30,2212;%%

%\cite{GarciaBellido:1999sv}
\bibitem{GarciaBellido:1999sv}
  J.~Garc\'{\i}a-Bellido, D.~Y.~Grigoriev, A.~Kusenko and M.~E.~Shaposhnikov,
  ``Non-equilibrium electroweak baryogenesis from preheating after
  inflation,''
  Phys.\ Rev.\  D {\bf 60}, 123504 (1999)
  [hep-ph/9902449].
  %%CITATION = PHRVA,D60,123504;%%



%\cite{Tranberg:2003gi}
\bibitem{Tranberg:2003gi}
  A.~Tranberg and J.~Smit,
  ``Baryon asymmetry from electroweak tachyonic preheating,''
  JHEP {\bf 0311} (2003) 016
  [hep-ph/0310342].
  %%CITATION = JHEPA,0311,016;%%


%\cite{Guth:1982pn}
\bibitem{Guth:1982pn}
  A.~H.~Guth and E.~J.~Weinberg,
  ``Could The Universe Have Recovered From A Slow First Order Phase
  Transition?,''
  Nucl.\ Phys.\  B {\bf 212}, 321 (1983).
  %%CITATION = NUPHA,B212,321;%%

%\cite{Nardini:2007me}
\bibitem{Nardini:2007me}
  G.~Nardini, M.~Quir\'os and A.~Wulzer,
  ``A Confining Strong First-Order Electroweak Phase Transition,''
  JHEP {\bf 0709} (2007) 077
  [hep-ph/0706.3388].
  %%CITATION = JHEPA,0709,077;%%

%\cite{Huber:2007vva}
\bibitem{Huber:2007vva}
  S.~J.~Huber and T.~Konstandin,
  ``Production of Gravitational Waves in the nMSSM,''
JCAP {\bf 0805} (2008) 017 
 [hep-ph/0709.2091].
  %%CITATION = ARXIV:0709.2091;%%


%\cite{Moore:1995ua}
\bibitem{Moore:1995ua}
  G.~D.~Moore and T.~Prokopec,
  ``Bubble wall velocity in a first order electroweak phase transition,''
  Phys.\ Rev.\ Lett.\  {\bf 75}, 777 (1995)
  [hep-ph/9503296].
  %%CITATION = PRLTA,75,777;%%

%\cite{Moore:1995si}
\bibitem{Moore:1995si}
  G.~D.~Moore and T.~Prokopec,
  ``How fast can the wall move? A Study of the electroweak phase transition
  dynamics,''
  Phys.\ Rev.\  D {\bf 52}, 7182 (1995)
  [hep-ph/9506475].
  %%CITATION = PHRVA,D52,7182;%%

%\cite{Moore:2000wx}
\bibitem{Moore:2000wx}
  G.~D.~Moore,
  ``Electroweak bubble wall friction: Analytic results,''
  JHEP {\bf 0003}, 006 (2000)
  [hep-ph/0001274].
  %%CITATION = JHEPA,0003,006;%%

%\cite{Steinhardt:1981ct}
\bibitem{Steinhardt:1981ct}
  P.~J.~Steinhardt,
  ``Relativistic Detonation Waves And Bubble Growth In False Vacuum Decay,''
  Phys.\ Rev.\  D {\bf 25} (1982) 2074.
  %%CITATION = PHRVA,D25,2074;%%

%\cite{Laine:1993ey}
\bibitem{Laine:1993ey}
  M.~Laine,
  ``Bubble Growth As A Detonation,''
  Phys.\ Rev.\  D {\bf 49} (1994) 3847
  [hep-ph/9309242].
  %%CITATION = PHRVA,D49,3847;%%

%\cite{Kamionkowski:1993fg}
\bibitem{Kamionkowski:1993fg}
  M.~Kamionkowski, A.~Kosowsky and M.~S.~Turner,
  ``Gravitational radiation from first order phase transitions,''
  Phys.\ Rev.\ D {\bf 49}, 2837 (1994)
  [astro-ph/9310044].
  %%CITATION = ASTRO-PH 9310044;%%

%\cite{Apreda:2001us}
\bibitem{Apreda:2001us}
R.~Apreda, M.~Maggiore, A.~Nicolis and A.~Riotto,
``Gravitational waves from electroweak phase transitions,''
Nucl.\ Phys.\ B {\bf 631} (2002) 342
[gr-qc/0107033].
%%CITATION = GR-QC 0107033;%%

%\cite{Caprini:2007xq}
\bibitem{Caprini:2007xq}
  C.~Caprini, R.~Durrer and G.~Servant,
  ``Gravitational wave generation from bubble collisions in first-order phase
  %transitions: an analytic approach,''
  Phys.\ Rev.\  D {\bf 77} (2008) 124015
  [astro-ph/0711.2593].
  %%CITATION = PHRVA,D77,124015;%%

%\cite{Huber:2008hg}
\bibitem{Huber:2008hg}
  S.~J.~Huber and T.~Konstandin,
  ``Gravitational Wave Production by Collisions: More Bubbles,''
  JCAP {\bf 0809}, 022 (2008)
  [arXiv:0806.1828 [hep-ph]].
  %%CITATION = JCAPA,0809,022;%%

%%%%%%%%%%%%%%%%%%%%%%%%%%%%%%%
\bibitem{GWturbulence}

%\cite{Kosowsky:2001xp}
%\bibitem{Kosowsky:2001xp}
  A.~Kosowsky, A.~Mack and T.~Kahniashvili,
  ``Gravitational radiation from cosmological turbulence,''
  Phys.\ Rev.\  D {\bf 66} (2002) 024030
  [astro-ph/0111483].
  %%CITATION = PHRVA,D66,024030;%%

%\cite{Dolgov:2002ra}
%\bibitem{Dolgov:2002ra}
  A.~D.~Dolgov, D.~Grasso and A.~Nicolis,
  ``Relic backgrounds of gravitational waves from cosmic turbulence,''
  Phys.\ Rev.\ D {\bf 66}, 103505 (2002)
  [astro-ph/0206461].
  %%CITATION = ASTRO-PH 0206461;%%

%\cite{Gogoberidze:2007an}
%\bibitem{Gogoberidze:2007an}
  G.~Gogoberidze, T.~Kahniashvili and A.~Kosowsky,
  ``The spectrum of gravitational radiation from primordial turbulence,''
  Phys.\ Rev.\  D {\bf 76} (2007) 083002
  [astro-ph/0705.1733].
  %%CITATION = PHRVA,D76,083002;%%

%\cite{Kahniashvili:2008er}
%\bibitem{Kahniashvili:2008er}
  T.~Kahniashvili, G.~Gogoberidze and B.~Ratra,
  ``Gravitational Radiation from Primordial Helical MHD Turbulence,''
  Phys.\ Rev.\ Lett.\  {\bf 100} (2008) 231301
  [astro-ph/0802.3524].
  %%CITATION = PRLTA,100,231301;%%

%\cite{Megevand:2008mg}
%\bibitem{Megevand:2008mg}
  A.~Megevand,
  ``Gravitational waves from deflagration bubbles in first-order phase
  transitions,''
  Phys.\ Rev.\  D {\bf 78} (2008) 084003
  [arXiv:0804.0391 [astro-ph]].
  %%CITATION = PHRVA,D78,084003;%%

%%%%%%%%%%%%%%%%%%%%%%%%%%%%%%%%%%
\bibitem{GWmag}

%\cite{Caprini:2006jb}
%\bibitem{Caprini:2006jb}
  C.~Caprini and R.~Durrer,
  ``Gravitational waves from stochastic relativistic sources: Primordial
  turbulence and magnetic fields,''
  Phys.\ Rev.\  D {\bf 74} (2006) 063521
  [astro-ph/0603476].
  %%CITATION = PHRVA,D74,063521;%%

%\cite{Fenu:2008vs}
%\bibitem{Fenu:2008vs}
  E.~Fenu and R.~Durrer,
  ``Interactions of cosmological gravity waves and magnetic fields,''
  [astro-ph/0809.1383].
  %%CITATION = ARXIV:0809.1383;%%


%%%%%%%%%%%%%%%%%%%%%%%%%%%%%%%%%%%%

%\cite{Nicolis:2003tg}
\bibitem{Nicolis:2003tg}
  A.~Nicolis,
  ``Relic gravitational waves from colliding bubbles and cosmic turbulence,''
  Class.\ Quant.\ Grav.\  {\bf 21} (2004) L27
  [gr-qc/0303084].
  %%CITATION = GR-QC 0303084;%%

%\cite{Grojean:2006bp}
\bibitem{Grojean:2006bp}
  C.~Grojean and G.~Servant,
  ``Gravitational waves from phase transitions at the electroweak scale and
  beyond,''
  Phys.\ Rev.\  D {\bf 75} (2007) 043507
  [hep-ph/0607107].
  %%CITATION = PHRVA,D75,043507;%%

%\cite{Silveira:1985rk}
\bibitem{Silveira:1985rk}
  V.~Silveira and A.~Zee,
  ``Scalar Phantoms,''
  Phys.\ Lett.\  B {\bf 161} (1985) 136.
  %%CITATION = PHLTA,B161,136;%%

%\cite{McDonald:1993ex}
\bibitem{McDonald:1993ex}
  J.~McDonald,
  ``Gauge Singlet Scalars as Cold Dark Matter,''
  Phys.\ Rev.\  D {\bf 50} (1994) 3637
  [hep-ph/0702143].
  %%CITATION = PHRVA,D50,3637;%%

%\cite{Burgess:2000yq}
\bibitem{Burgess:2000yq}
  C.~P.~Burgess, M.~Pospelov and T.~ter Veldhuis,
  ``The minimal model of nonbaryonic dark matter: A singlet scalar,''
  Nucl.\ Phys.\  B {\bf 619} (2001) 709
  [hep-ph/0011335].
  %%CITATION = NUPHA,B619,709;%%

%\cite{Hambye:2007vf}
\bibitem{Hambye:2007vf}
  T.~Hambye and M.~H.~G.~Tytgat,
  ``Electroweak Symmetry Breaking induced by Dark Matter,''
  Phys.\ Lett.\  B {\bf 659} (2008) 651
  [arXiv:0707.0633 [hep-ph]].
  %%CITATION = PHLTA,B659,651;%%

%\cite{Lee:1977ua}
\bibitem{Lee:1977ua}
  B.~W.~Lee and S.~Weinberg,
  ``Cosmological lower bound on heavy-neutrino masses,''
  Phys.\ Rev.\ Lett.\  {\bf 39} (1977) 165.
  %%CITATION = PRLTA,39,165;%%

%\cite{Bodeker:2004ws}
\bibitem{Bodeker:2004ws}
  D.~Bodeker, L.~Fromme, S.~J.~Huber and M.~Seniuch,
  ``The baryon asymmetry in the standard model with a low cut-off,''
  JHEP {\bf 0502}, 026 (2005)
  [arXiv:hep-ph/0412366].
  %%CITATION = JHEPA,0502,026;%%

\end{thebibliography}
\end{document}